\documentclass[a4paper,11pt]{article}

\usepackage{jinstpub}
\usepackage{verbatim}
\usepackage[compact]{titlesec}

\usepackage{lineno}

\title{\boldmath Neutral Bremsstrahlung in TPCs}

\author[a]{P. Amedo}
\author[a]{, D. González-Díaz}
\author[b]{, B. J. P. Jones}

\affiliation[a]{Instituto Gallego de Física de Altas Energías, Univ. de Santiago de Compostela, Campus sur, Rúa Xosé María Suárez Núñez, s/n, Santiago de Compostela, E-15782, Spain}
\affiliation[b]{Department of Physics, University of Texas at Arlington, Arlington, TX 76019, USA}

\emailAdd{pablo.amedo.martinez@usc.es}

\abstract{Traditionally, it has been assumed that electroluminescence (EL) in time projection chambers was purely an excimer-based emission. This idea changed when neutral bremsstrahlung (NBrS) was observed first in argon and subsequently in xenon a few years ago. In this work we explore further the framework used to explain these observations, presenting results for noble gas -based mixtures, as well as mixtures including small fractions of a molecular additive (`quencher'). Spectral content and yields are discussed in some cases of contemporary interest, together with their pressure-scalings.}


\proceeding{LIDINE 2021: Conference in LIght Detection of Noble Elements\\
  14-17 September 2021\\
  San Diego, California, USA}

\begin{document}
\maketitle
\flushbottom


\titlespacing{\section}{0pt}{1ex}{0ex}
\setlength{\abovecaptionskip}{4pt}
\setlength{\belowcaptionskip}{-10pt}
\section{Introduction}

Since their introduction in 1974 \cite{Nygren}, time projection chambers (TPCs) have proved to be one of the most effective ways of detecting particles and reconstructing their trajectories. The versatility of these devices, being compatible with $B$-fields, allowing readout flexibility and a wide range of density media (from some 10's of mbar up to liquid or even solid phase), makes them the perfect tool to study many different phenomena in particle physics \cite{DiegoReview}.

In particular, TPCs making use of their scintillation signal (either primary or secondary) are being used in different experiments, such as direct dark matter searches \cite{1,2}, neutrinoless double-beta decay ($\beta\beta0\nu$) \cite{3} or double electron capture \cite{4}, but also in future neutrino experiments such as DUNE \cite{5}. Depending on which noble element is used, they usually rely on well-known scintillation bands, such as the “second continuum” bands in xenon or argon, which are around 172~nm  and 128~nm in gas, respectively \cite{Takahashi}. 

Until very recently, it was believed that secondary scintillation in noble element TPCs was produced solely from the excited states created by the primary ionization electrons when drifting across a high field region (conveniently situated at the anode plane). These states result largely on excimer production through three body processes, giving rise to those well-known emission bands \cite{AprileBook}. This notion was challenged three years ago \cite{Buzu} by showing that there was light emission below the classical electroluminescence (EL) threshold of argon and proposing a mechanism to explain it: neutral bremsstrahlung (NBrS). Since then further studies were conducted, showing its presence in Xe and Xe-C$_2$H$_6$ mixtures and helping to firmly establish the phenomenon \cite{PRX, TalkLIDINE}.

Regarding the potential impact in real-world experiments, the readout of a dual-phase TPC demonstrator in the visible-band by making a partial use of this emission was performed in \cite{DarSide_NBrS}. Despite the yields being much lower than those achieved through excimer emission, a number of advantages like the immunity of NBrS to quenching by impurities, fast nature, broad-band characteristics and low electric fields needed for operation make it a subject of ongoing technological interest. At the same time, the phenomenon has been shown to have an impact in ongoing experiments, causing background scintillation in  the 'buffer' regions of single-drift (asymmetric) TPCs, i.e. between the cathode and photosensor plane \cite{PRX}.

In this work we explore the possibilities of the framework used to calculate NBrS in \cite{PRX}. First we will introduce briefly the theoretical foundations of NBrS and explain how the calculations can be performed, giving results for a number of weakly quenched noble gases, conditions under which the current framework is expected to apply. Finally we will discuss the density scalings of the NBrS yields and its verification in transport.

\section{Theory and calculation}

In a process analogous to nuclear bremsstrahlung, NBrS can be viewed as the interaction of an electron with the dipole field of a neutral atom thereby leading to the emission of a photon, while conserving energy and momentum thanks to the atom recoil \cite{Dalgarno,Johnston,Ohmura,Geltman}. This is possible even in the absence of permanent dipole moments of the gas species, as those can still interact with the incoming electron via induced dipole moments.

In order to study NBrS, Fermi's golden rule can be used to calculate the transition probability between the initial/final state of the incoming/scattered electron. This leads to the following result for the differential cross section per unit of photon frequency \cite{Dalgarno}:

\begin{equation}
\frac{d\sigma}{d\nu} = \frac{8 \pi e^2 \nu^3 m_e^2 k_f}{3 \hbar^3 c^3 k_i} |M|^2
\label{FermiG}
\end{equation}
being ${\hbar}k_{i(f)}$ the initial (final) momentum of the electron and $M$ a matrix element involving the initial and final states:

\begin{equation}
|M|^2 \equiv |\langle \Psi_f|\vec{r}|\Psi_i \rangle|^2 \label{matrix1}
\end{equation}
Those may be approximated by using a scattering formalism based on partial waves and keeping the first terms. If approximating those by the ones used for computations in the case of the hydrogen ion (with $1\%$ accuracy in that case, according to \cite{Ohmura}), the following expression for the matrix element can be derived \cite{Dalgarno}:

\begin{equation}
|M|^2 = \frac{64\pi^2}{(k_i^2 - k_f^2)^4} [k_i^2 Q(k_f) + k_f^2 Q(k_i)] \label{Mcalc}
\end{equation}
thus displaying explicit proportionality with the elastic scattering cross section $Q$. By plugging this differential cross section directly into equation \ref{FermiG} and recalling the relationships $\varepsilon_{i,f} = (\hbar^2/2m_e)k_{i,f}^2$, $h\nu = \varepsilon_i - \varepsilon_f$, we can obtain the final expression for the cross section per unit of photon frequency:

\begin{equation}
\frac{d\sigma}{d\nu} = \frac{8}{3}\frac{r_e}{c}\frac{1}{h\nu}\left(\frac{\varepsilon_i-h\nu}{\varepsilon_i}\right)^{1/2} \cdot \left[\varepsilon_i \cdot{Q_{(m)}}(\varepsilon_i-h\nu) + (\varepsilon_i - h\nu)\cdot{Q_{(m)}}(\varepsilon_i) \right]\label{nBr_eq4}
\end{equation}
Here we make explicit an ambiguity in the existing analytical derivations, through which the elastic and momentum transfer (subindex $m$) cross sections can be used interchangeably (see \cite{PRX} for a discussion). In the following we will assume proportionality with $Q_m$ for simplicity, although the differences are generally small and at the level of 10's of a $\%$.

In order to compute the NBrS emission rate in a TPC, a swarm of ionization electrons needs to be considered, characterized through an energy distribution $dP/d\varepsilon$. This is done by simply averaging the product of the energy-dependent velocity of the electrons ($v(\epsilon)$), and the differential cross section for emission ($d\sigma/d\nu$) over the energy distribution of the electrons:

\begin{equation}
\frac{dN_\gamma}{d\nu dt} = \int_0^{\infty} N \frac{d\sigma}{d\nu} v(\varepsilon) \frac{dP}{d\varepsilon} d{\varepsilon} \label{nBr_eq2}
\end{equation}
\begin{figure}[h!!!]
\centering 
\includegraphics[width=0.9\textwidth,origin=c,angle=0]{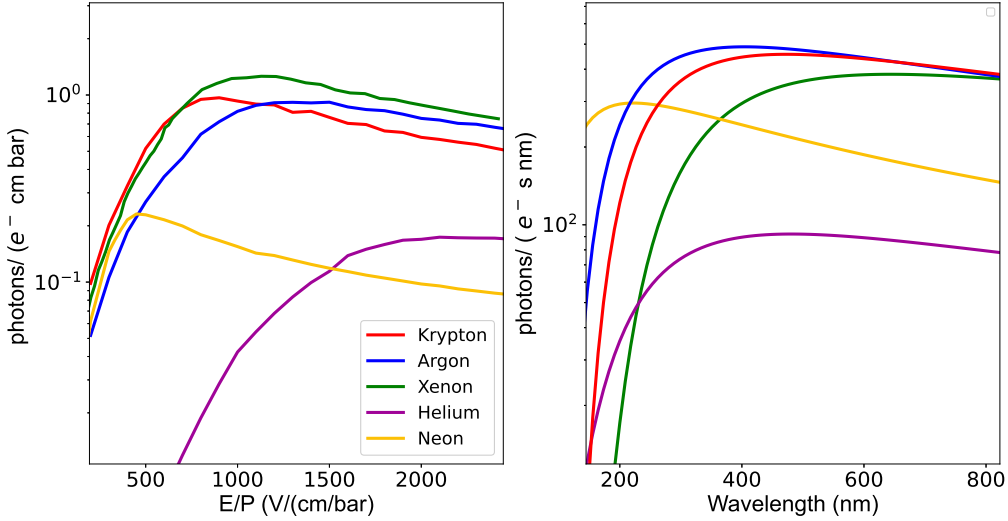}
\caption{\label{fig:spectrum} Left: pressure-reduced scintillation yields for NBrS in noble gases as a function of electric field (integrated over the practical detection range 120-1000~nm). Right: example of emission spectra (rate per electron, second and nanometer) at a pressure-reduced electric field of $E/P=1200$ V/cm/bar, $T=300$~K and $P=1$~bar. }
\end{figure}

The yield per unit length, a magnitude easier to determine experimentally than the scintillation rate, can be calculated in the following way:
\begin{equation}
\label{nBr_eq6}
Y \equiv \frac{dN_\gamma}{dz} = \frac{1}{v_d} \int_{\nu_{min}}^{\nu_{max}} \frac{dN_\gamma}{d\nu dt} = \frac{1}{v_d} \int_{\nu_{min}}^{\nu_{max}}\int_0^{\infty} N \frac{d\sigma}{d\nu} v(\varepsilon) \frac{dP}{d\varepsilon} d{\varepsilon} d{\nu}
\end{equation}
where $v_d$ is the drift velocity of the electrons in the gas. The above formalism can be easily extended for a weakly quenched gas (meaning in this context that NBrS emission involving inelastic degrees of freedom is expected to be subdominant). The yield calculation can be performed in the following way for a two-component mixture with concentrations $f_{A}$ and $f_{B}$, and cross sections $\sigma_{A}$ and $\sigma_{B}$:\footnote{An arbitrary number of compounds can be considered, but for simplicity we discuss only binary mixtures in this work.}

\begin{equation}
\label{nBr_mixt}
Y_{AB} \left(dP/d\varepsilon \right) =f_A \cdot Y_{A}\left( dP/d\varepsilon, \sigma_{A}\right)
+ (1-f_A) \cdot Y_{B}\left( dP/d\varepsilon, \sigma_{B}\right)
\end{equation}
In order to have access to both the drift velocity and energy distributions, Monte Carlo simulations with a locally modified version of the Pyboltz transport code \cite{Pyboltz} have been realized.

\section{Spectrum}

NBrS is a broadband emission with a blue-wing depending on the maximum electron energy (equivalently, on the density-reduced electric field). Spectral emission rates obtained for noble gases based on expression \ref{nBr_eq2} can be found in figure \ref{fig:spectrum}. A reduced field roughly corresponding to the maximum of the emission was chosen.

\section{Noble and weakly-quenched gas mixtures}

\begin{figure}[htbp]
\centering 
\includegraphics[width=1\textwidth,origin=c,angle=0]{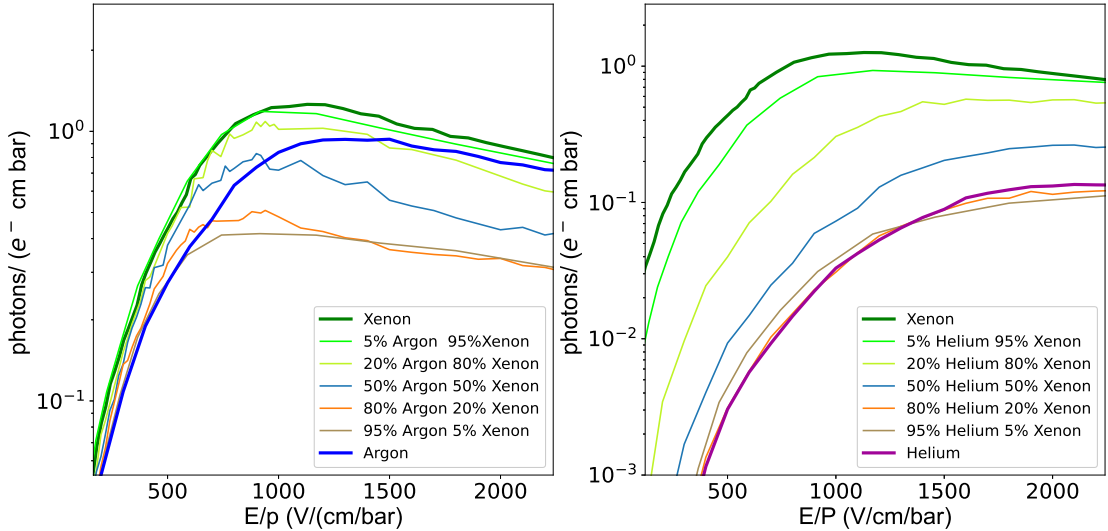}
\caption{\label{fig:yields} NBrS yield for mixtures based on xenon-argon (left) and xenon-helium (right), integrated over the wavelength range from 120 to 1000~nm, at $T=300~$K and $P=1~$bar.}
\end{figure}

The integrated yields of NBrS (considered here in the practical range of interest of 120-1000~nm) can be `tuned' to a considerable extent in noble element-mixtures, through the interplay between the different position of the Ramsauer minima, the impact of the cross sections on the energy distribution of the electrons (e.g., through the higher cooling power of helium) as well as the overall magnitude of the cross sections, that are directly proportional to the emission probability (e.g., with xenon cross section being generally higher at low energies). Figure \ref{fig:yields} shows for illustration tbe NBrS yields in mixtures of xenon-argon (left) and xenon-helium (right). In particular, as Xe-He mixtures are being considered for electroluminescence-based time projection chambers due to their low electron diffusion \cite{XeHeTPCs}, it is interesting to note that the spurious presence of NBrS in the chamber would be beneficially suppressed.

\begin{figure}[h!!!]
\centering 
\includegraphics[width=.7\textwidth,origin=c,angle=0]{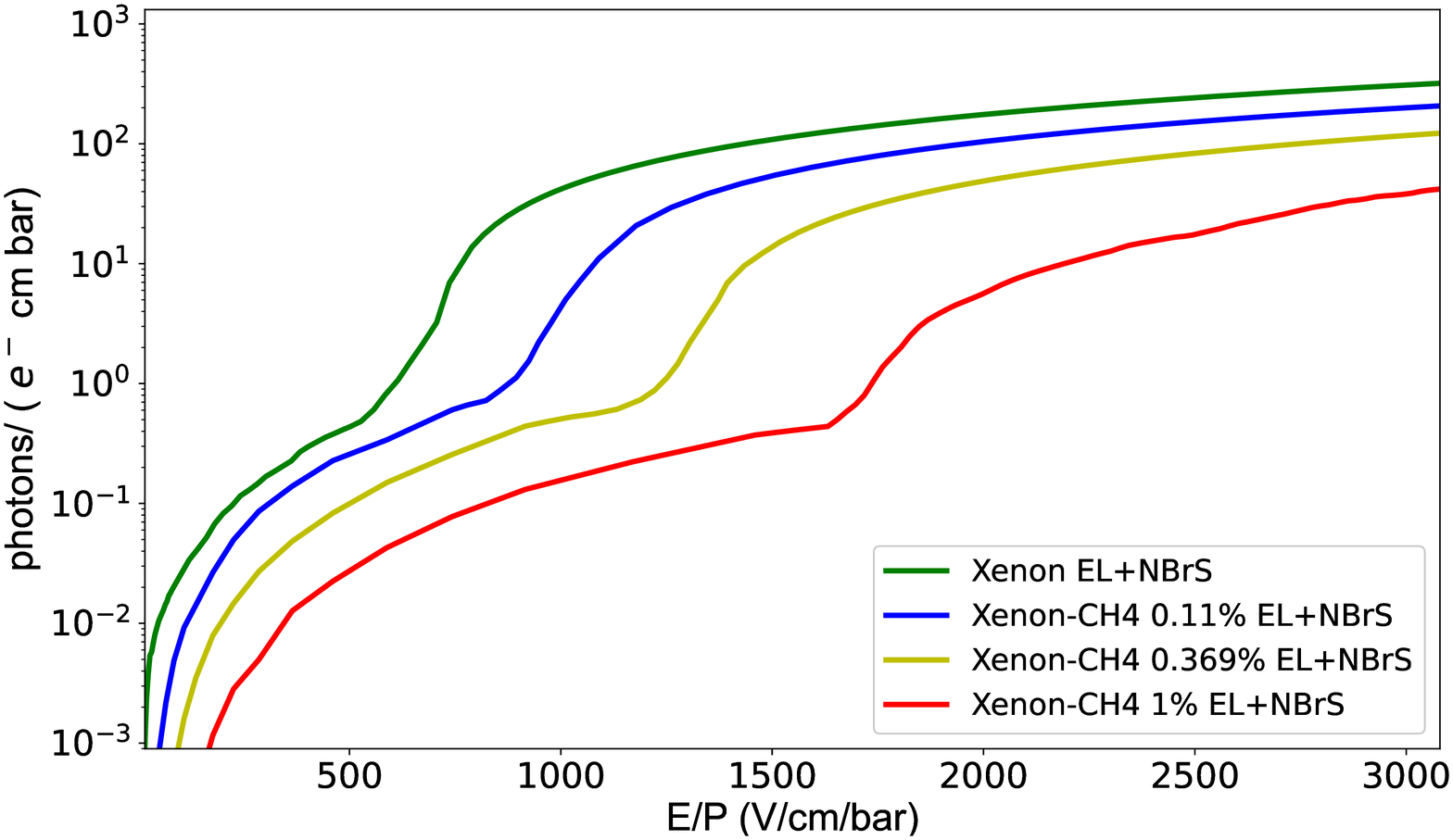}
\caption{\label{fig:CH4} Scintillation yields of Xe-CH$_4$ mixtures as a function of the pressure-reduced electric field (integrated over the practical detection range 120-1000~nm). The chosen dopings are $0.11\%$, $0.37\%$ and $1\%$ (molar), and the gas operating conditions are 300~K and 1~bar. EL simulations have been taken from \cite{CH4yields}.}
\end{figure}

Weakly quenched gas mixtures can be simulated as well, as it is seen in figure \ref{fig:CH4} where Xe-CH$_4$ has been chosen for the purpose of illustration. In this case, the EL simulations from \cite{CH4yields} have been added to illustrate the contributions of both electron cooling (shifting the curves towards higher $E/P$ values) and scintillation quenching (suppressing the yields in the EL region) to the overall gas scintillation.

\section{Density scalings}

\begin{figure}[h!!!]
\centering 
\includegraphics[width=.8\textwidth,origin=c,angle=0]{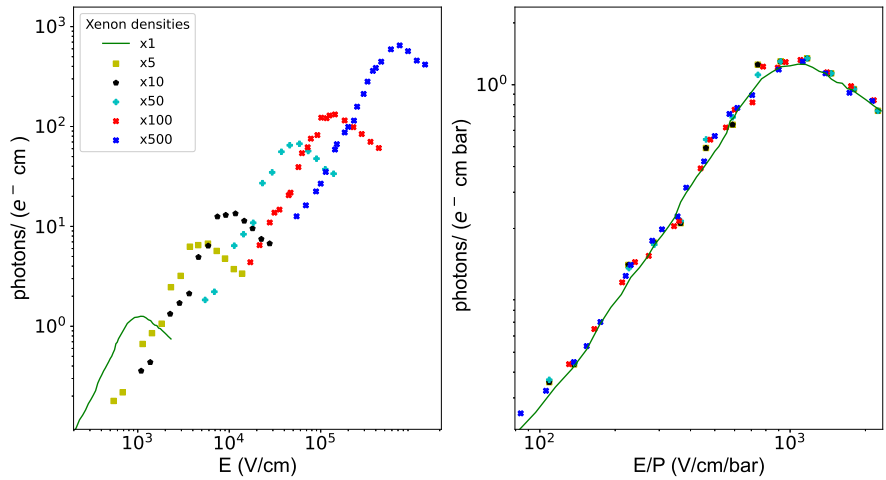}
\caption{\label{fig:scaling} NBrS yield calculated for different densities of gaseous xenon in the range of 120 to 1000 nm at $T=300$~K. The green line corresponds to a reference simulation performed for $P=1$~bar and with a large number of collisions. At higher fields and densities the absolute yield increases (left plot) but the reduced yield stays the same at the same reduced electric field (right plot).}
\end{figure}

NBrS is expected to show scalings with the number density, $N$, (or pressure, at fixed temperature). The $N$-scaling on the yield, $Y$, can be understood from the fact that the probability of NBrS emission per collision is constant, so the yields are bound to depend linearly on the gas gap and pressure. Concerning the electric field, as long as the energy distribution depends on its density-reduced value ($E/N$), $Y/N$ can be expected to be a function of $E/N$ too. The validity of these scalings in Monte Carlo is demonstrated over almost three orders of magnitude in figure \ref{fig:scaling}.

\section{Conclusions}

In this paper we have shown how to easily perform NBrS calculations for a wide variety of mixtures and elements, resorting to the open-source code Pyboltz. The predictive capability anticipated from the data-benchmarking performed with xenon in \cite{PRX}, should allow future experiments to take informed decisions on regard to this new effect. Implementation of the above algorithms in the Pyboltz repository \cite{PyBoltzGithub} is currently being undertaken and expected to happen within weeks of the time of publication of this work.


\end{document}